\documentclass[reprint,amsmath,amssymb]{revtex4-1}
\usepackage[utf8]{inputenc}
\usepackage[T1]{fontenc}
\usepackage{graphicx}
\usepackage{dcolumn}
\usepackage{bm}
\usepackage{enumerate}

\begin{document}

\preprint{}

\title{Connection independent formulation of general relativity}

\author{Junpei Harada}
 \email{jharada@hoku-iryo-u.ac.jp}
\affiliation{Health Sciences University of Hokkaido, Japan}

\date{February 4, 2020}

\begin{abstract}
A connection-independent formulation of general relativity is presented, in which the dynamics does not depend on the choice of connection. The gravity action in this formulation includes one additional scalar term in addition to the Einstein-Hilbert action. No conditions on the connection are imposed. Nevertheless, this formulation yields the Einstein equations, without adding the Gibbons-Hawking-York term even when a manifold has a boundary. Furthermore, this formulation yields a unified description of general relativity, teleparallel gravity, and symmetric teleparallel gravity. 
\end{abstract}
\maketitle

\section{Introduction}
On any manifold, there are infinitely many affine connections to define the covariant derivative, but {\it a priori}, no one is better than the others. A choice of connections is arbitrary, and hence it should play no role in the formulation of any physical law including gravity. 

However, in the conventional formulations of gravity theories, a specific connection has been used by imposing some conditions on the connection. For instance, the original formulation of general relativity uses the Levi-Civita connection by imposing two conditions, both metricity and torsion-free. In the Palatini-Einstein formalism~\cite{Ferraris:1982vfg}, where the connection and the metric are taken as independent variables, a specific connection is not adopted. However, even in that formalism, either metricity or torsion-free is imposed, and then varying the action with respect to the connection, one learns that the connection coincides with the Levi-Civita connection. 

In this paper the connection-independent formulation of the dynamics of general relativity is presented. No conditions on the connection are imposed. This formulation necessarily introduces one additional scalar term in addition to the Einstein-Hilbert action. The variation of the action with respect to the connection identically vanishes, and the variation of the action with respect to the metric yields the Einstein equations. It is not necessary to add the Gibbons-Hawking-York term~\cite{Gibbons:1976ue,York:1972sj} even when a manifold has a boundary. Furthermore, the action in this formulation yields general relativity, teleparallel gravity, symmetric teleparallel gravity, and others. 

The paper is organized as follows. In section~\ref{sec:ci} the gravity action is given, which defines the connection-independent formulation. In section~\ref{sec:variation} the variation of the action is investigated. In section~\ref{sec:GHY} it is shown that the action includes the Gibbons-Hawking-York term. In section~\ref{sec:formulations}, the original formulation of general relativity, the Palatini-Einstein formalism, and the formulation presented in this paper are summarized for comparison. In section~\ref{sec:examples}, it is demonstrated that the action yields general relativity, teleparallel gravity, symmetric teleparallel gravity, and others. Section~\ref{sec:conclusion} is devoted to conclusion.

\section{Connection independence\label{sec:ci}}
Consider a vector field on a spacetime manifold $M$. We denote the covariant derivative of a vector $V_\nu$ as
\begin{align}
	\nabla_\mu V_\nu =\partial_\mu V_\nu - \Gamma^\lambda{}_{\mu\nu}V_\lambda.
	\label{def:cderivative}
\end{align}
Eq.~\eqref{def:cderivative} remains a tensor if one adds any tensor $\Omega^\lambda{}_{\mu\nu}$ to the affine connection $\Gamma^\lambda{}_{\mu\nu}$,
\begin{align}
	\Gamma^\lambda{}_{\mu\nu} \rightarrow \Gamma^\lambda{}_{\mu\nu} + \Omega^\lambda{}_{\mu\nu}.
	\label{trans:affine}
\end{align}
Each choice of $\Omega^\lambda{}_{\mu\nu}$ defines different affine connections. This implies that there are infinitely many affine connections, because a tensor $\Omega^\lambda{}_{\mu\nu}$ is arbitrary.

Of infinitely many possible affine connections, general relativity uses a specific one, the Levi-Civita connection
\begin{align}
	\overline\Gamma^\lambda{}_{\mu\nu}
	\equiv
	\frac{1}{2}g^{\lambda\rho} \left(\partial_\mu g_{\nu\rho}+\partial_\nu g_{\mu\rho} - \partial_\rho g_{\mu\nu}\right). 
	\label{def:Levi-Civita}
\end{align}
This leads to nonzero spacetime curvature but zero torsion. This is also compatible with the metric, $\overline{\nabla}_\lambda g_{\mu\nu}=0$.
A bar is used to denote the quantities associated with the Levi-Civita connection. It is convenient to use the Levi-Civita connection as the reference for other connections. Then, an arbitrary connection $\Gamma^\lambda{}_{\mu\nu}$ can be written as
\begin{align}
	\Gamma^\lambda{}_{\mu\nu} 
	= \overline\Gamma^\lambda{}_{\mu\nu} + W^\lambda{}_{\mu\nu},
	\label{def:tensorW}
\end{align}
where $W^\lambda{}_{\mu\nu}$ is a tensor, because the difference between two affine connections is a tensor.

The Riemann curvature tensor is defined by
\begin{align}
	R^\rho{}_{\mu\lambda\nu}
	= \partial_\lambda \Gamma^\rho{}_{\nu\mu}
	 - \partial_\nu \Gamma^\rho{}_{\lambda\mu}
	 + \Gamma^\rho{}_{\lambda\sigma} \Gamma^\sigma{}_{\nu\mu}
	 -  \Gamma^\rho{}_{\nu\sigma} \Gamma^\sigma{}_{\lambda\mu},
\end{align}
and this transforms under the transformation~\eqref{trans:affine} as 
\begin{align}
	&R^\rho{}_{\mu\lambda\nu} 
	\rightarrow 
	R^\rho{}_{\mu\lambda\nu} 
	+ \Omega^\rho{}_{\lambda\sigma}\Omega^\sigma{}_{\nu\mu} 
	- \Omega^\rho{}_{\nu\sigma}\Omega^\sigma{}_{\lambda\mu} \nonumber\\
	&+W^\rho{}_{\lambda\sigma}\Omega^\sigma{}_{\nu\mu} 
	- W^\sigma{}_{\lambda\mu}\Omega^\rho{}_{\nu\sigma}
	- W^\rho{}_{\nu\sigma}\Omega^\sigma{}_{\lambda\mu} 
	+ W^\sigma{}_{\nu\mu}\Omega^\rho{}_{\lambda\sigma}\nonumber\\
	&+\overline{\nabla}_\lambda \Omega^\rho{}_{\nu\mu} 
	- \overline{\nabla}_\nu \Omega^\rho{}_{\lambda\mu},
\end{align}
where $\overline{\nabla}_\mu$ is the covariant derivative with respect to the Levi-Civita connection \eqref{def:Levi-Civita}, rather than an arbitrary one. Then, the Ricci scalar $R\equiv g^{\mu\nu}R^\lambda{}_{\mu\lambda\nu}$ transforms under the transformation~\eqref{trans:affine} as 
\begin{align}
	R& 
	\rightarrow R
	+ \Omega^\lambda{}_{\lambda\nu}\Omega^{\nu\mu}{}_{\mu} 
	- \Omega^{\mu\nu}{}_{\lambda}\Omega^\lambda{}_{\mu\nu} \nonumber\\
	+&W^\lambda{}_{\lambda\nu}\Omega^{\nu\mu}{}_{\mu} 
	- W^\lambda{}_{\mu\nu}\Omega^{\mu\nu}{}_{\lambda}
	- W^{\mu\nu}{}_{\lambda}\Omega^\lambda{}_{\mu\nu} 
	+ W^{\nu\mu}{}_{\mu}\Omega^\lambda{}_{\lambda\nu}\nonumber\\
	+&\overline{\nabla}_\mu \left(\Omega^{\mu\nu}{}_{\nu} - \Omega_\nu{}^{\nu\mu}\right).
	\label{trans:Ricci}
\end{align}
This indicates that the Einstein-Hilbert action
\begin{align} 
	S_{\rm EH} = \frac{1}{16\pi} \int_M d^4 x \sqrt{-g} R
	\label{action:EH}
\end{align}
is not invariant under the transformation~\eqref{trans:affine}. 

We now want to find a gravity action which is invariant under the transformation~\eqref{trans:affine}. For this purpose, the following scalar $W$ should be defined,
\begin{align} 
	W \equiv W^{\mu\nu}{}_\lambda W^\lambda{}_{\mu\nu} - W^\lambda{}_{\lambda\nu} W^{\nu\mu}{}_\mu.
	\label{def:scalarW}
\end{align}
Under the transformation~\eqref{trans:affine}, the $W^\lambda{}_{\mu\nu}$ transforms as $W^\lambda{}_{\mu\nu} \rightarrow W^\lambda{}_{\mu\nu} + \Omega^\lambda{}_{\mu\nu}$, and then the $W$ transforms as
\begin{align}
	W 
	&\rightarrow 
	W
	+\Omega^{\mu\nu}{}_{\lambda}\Omega^\lambda{}_{\mu\nu} 
	- \Omega^\lambda{}_{\lambda\nu}\Omega^{\nu\mu}{}_{\mu}\nonumber\\
	+&W^{\mu\nu}{}_{\lambda}\Omega^\lambda{}_{\mu\nu} 
	+ W^\lambda{}_{\mu\nu}\Omega^{\mu\nu}{}_{\lambda}
	-W^\lambda{}_{\lambda\nu}\Omega^{\nu\mu}{}_{\mu} 	
	- W^{\nu\mu}{}_{\mu}\Omega^\lambda{}_{\lambda\nu}.
	\label{trans:scalarW}	
\end{align} 
Thus, neither the Ricci scalar $R$ nor the scalar $W$ is invariant under the transformation~\eqref{trans:affine}.

However, we learn from~\eqref{trans:Ricci} and~\eqref{trans:scalarW} that the $R+W$ transforms under the transformation~\eqref{trans:affine} as
\begin{align}
	R+W 
	&\rightarrow 
	R+W
	+\overline{\nabla}_\mu \left(\Omega^{\mu\nu}{}_{\nu} - \Omega_\nu{}^{\nu\mu}\right).
	\label{trans:R+W}
\end{align}
This indicates that the $R+W$ is invariant under the transformation~\eqref{trans:affine} up to the divergence term. 
The transformation law~\eqref{trans:R+W} implies that the following action
 \begin{align} 
	S = \frac{1}{16\pi } \int_M d^4 x \sqrt{-g} (R+W)	
	\label{action:R+W}
\end{align}
defines a connection-independent formulation of gravity. We can take \eqref{action:R+W} as the gravity action rather than \eqref{action:EH}.

\section{Variation of the action\label{sec:variation}}
We investigate the variation of the action~\eqref{action:R+W} with respect to the connection and the metric independently. It is convenient to rewrite~\eqref{action:R+W}, instead of varying~\eqref{action:R+W} itself. Using \eqref{trans:R+W}, we can write the $R+W$ as
\begin{align}
	R + W 
	= \overline{R} 
	+ \overline{\nabla}_\mu \left(W^{\mu\nu}{}_\nu - W_\nu{}^{\nu\mu}\right),
	\label{R+W:tensorW}
\end{align}
where $\overline{R}$ is the Ricci scalar of the Levi-Civita connection. This identity is useful, because it is valid for any connection. It indicates that the $R+W$ differs from $\overline{R}$ in only the divergence term.
Substituting \eqref{R+W:tensorW} into \eqref{action:R+W}, we obtain
\begin{align}
	S&=
	\frac{1}{16\pi} \int_M d^4 x \sqrt{-g} \  \overline{R}\nonumber\\
	&+\frac{1}{16\pi} \int_M d^4 x \sqrt{-g} \  \overline{\nabla}_\mu \left(W^{\mu\nu}{}_\nu - W_\nu{}^{\nu\mu}\right),
	\label{action:GR+boundary}	
\end{align}
where
\begin{align}
	&W^{\mu\nu}{}_\nu - W_\nu{}^{\nu\mu}\nonumber\\
	&=
	g^{\rho\sigma}\Gamma^\mu{}_{\rho\sigma}-g^{\mu\nu}\Gamma^\lambda{}_{\lambda\nu}
	-g^{\rho\sigma}\overline\Gamma^\mu{}_{\rho\sigma} + g^{\mu\nu}\overline\Gamma^\lambda{}_{\lambda\nu}.
	\label{R+W:divergence}
\end{align}

Varying~\eqref{action:GR+boundary} with respect to the connection, we immediately find that the variation $\delta_\Gamma S$ identically vanishes, 
\begin{align}
	\delta_\Gamma S = 0,
	\label{variation:connection}
\end{align}
under the boundary condition $\delta\Gamma^\lambda{}_{\mu\nu}=0$. This implies that the connection $\Gamma$ is not a dynamical variable.

Varying \eqref{action:GR+boundary} with respect to the metric, we can write
\begin{align}
	\delta_g S = \delta_g S_{\rm GR}+\delta_g S_{\rm B}, 
\end{align}
where $S_{\rm GR}$ and $S_{\rm B}$ denote the first term and the boundary term in~\eqref{action:GR+boundary}, respectively. As usual, $\delta_g S_{\rm GR}$ is
 \begin{align} 
	\delta_g S_{\rm GR}
	&= \frac{1}{16\pi } \int_M d^4 x \sqrt{-g} \left(\overline{R}_{\mu\nu} - \frac{1}{2}\overline{R} g_{\mu\nu}\right)\delta g^{\mu\nu}	\nonumber\\
	&+\frac{1}{16\pi } \int_M d^4 x \sqrt{-g} g^{\mu\nu} \delta \overline{R}_{\mu\nu},
	\label{variation:GR}
\end{align}
where $\delta g_{\mu\nu} = \delta g = 0$ on the boundary to be understood. The second term in~\eqref{variation:GR} is nonzero when a manifold has a boundary~\cite{York:1972sj}.
Then, the variation $\delta_g S_{\rm B}$ is obtained as
\begin{align} 
	\delta_g S_{\rm B}
	&=
	 -\frac{1}{16\pi} \int_M d^4 x \sqrt{-g} \  \overline{\nabla}_\mu
	 \left(g^{\rho\sigma}\delta\overline\Gamma^\mu{}_{\rho\sigma} 
	 - g^{\mu\nu}\delta\overline\Gamma^\lambda{}_{\lambda\nu}\right)\nonumber\\
	 &=
	 - \frac{1}{16\pi } \int_M d^4 x \sqrt{-g} g^{\mu\nu} \delta \overline{R}_{\mu\nu},
	\label{variation:boundary}
\end{align}
where we have used the Palatini identity,
\begin{align}
	\delta \overline{R}_{\mu\nu}
	= \overline{\nabla}_\lambda \delta \overline\Gamma^\lambda{}_{\mu\nu}
	-\overline{\nabla}_\nu\delta \overline\Gamma^\lambda{}_{\lambda\mu}.
	\label{identity:Palatini}
\end{align}
Thus, \eqref{variation:boundary} exactly cancels the second term in~\eqref{variation:GR}. As the result, the variation $\delta_g S$ is given by 
 \begin{align} 
	\delta_g S 
	=& \frac{1}{16\pi } \int_M d^4 x \sqrt{-g} \left(\overline{R}_{\mu\nu} - \frac{1}{2}\overline{R} g_{\mu\nu}\right)\delta g^{\mu\nu}.
	\label{variation:gravity}
\end{align}
Here, it should be noted that we have not introduced the Gibbons-Hawking-York term to obtain~\eqref{variation:gravity}.

\section{The Gibbons-Hawking-York term \label{sec:GHY}}
In general relativity, the usual procedure to cancel the second term in~\eqref{variation:GR} is to add the Gibbons-Hawking-York term~\cite{Gibbons:1976ue,York:1972sj} in addition to the Einstein-Hilbert action. However, in section~\ref{sec:variation}, we have shown that~\eqref{variation:boundary} cancels the second term in~\eqref{variation:GR}. It means that~\eqref{variation:boundary} plays the role of the variation of the Gibbons-Hawking-York term. Therefore, in the present formulation, there is no need to add the Gibbons-Hawking-York term. Furthermore, we can show that~\eqref{action:R+W} includes the Gibbons-Hawking-York term itself. This can be shown as follows.

The Gibbons-Hawking-York term is defined by
\begin{align}
	S_{\rm GHY} \equiv \frac{\epsilon}{8\pi} \oint_{\partial M} d^3 y \sqrt{|h|} \ \overline{\nabla}_\mu n^\mu,
	\label{action:GHY}
\end{align}
where $n_\mu$ is a unit vector normal to the boundary, $h$ is the determinant of the induced metric on the boundary, $y$ are the coordinates on the boundary, $\epsilon \equiv n^\mu n_\mu$ is equal to +1 where $\partial M$ is timelike and $-1$ where $\partial M$ is spacelike, and $ \overline{\nabla}_\mu n^\mu \equiv K$ is the trace of the extrinsic curvature. We show that $S_{\rm B}$ includes $S_{\rm GHY}$ in the following.

The dynamical term in $S_{\rm B}$ is given by
\begin{align}
	\frac{1}{16\pi} \int_M d^4 x \sqrt{-g} \  \overline{\nabla}_\mu 
	\left(-g^{\rho\sigma}\overline\Gamma^\mu{}_{\rho\sigma} 
	+ g^{\mu\nu}\overline\Gamma^\lambda{}_{\lambda\nu}\right),
	\label{action:dynamical_boundary}
\end{align}
where the non-dynamical term including $\Gamma$ is omitted, because it does not affect the variation and field equations. Using the Stokes' theorem, we learn that \eqref{action:dynamical_boundary} equals to 
\begin{align}
	&\frac{\epsilon}{16\pi} \oint_{\partial M} d^3 y \sqrt{|h|} \  
	n_\mu 
	\left(-g^{\rho\sigma}\overline\Gamma^\mu{}_{\rho\sigma} 
	+ g^{\mu\nu}\overline\Gamma^\lambda{}_{\lambda\nu}\right)\nonumber\\
	&=\frac{\epsilon}{8\pi} \oint_{\partial M} d^3 y \sqrt{|h|} \ \overline{\nabla}_\mu n^\mu.
	\label{ointegration:dynamical_boundary}
\end{align}
This is identical to the Gibbons-Hawking-York term~\eqref{action:GHY}. Thus, in our formulation, it is not necessary to add~$S_{\rm GHY}$, because the action~\eqref{action:R+W} includes it. 

\section{Three formulations\label{sec:formulations}}
It would be useful to summarize the three formulations, the original formulation of general relativity, the Palatini-Einstein formalism, and the formulation of this paper. 

\begin{enumerate}
	\item {\it Original formulation of general relativity}: The action is defined by the first term in~\eqref{action:GR+boundary}. The Levi-Civita connection is used by imposing two conditions on the connection, metricity and torsion-free. When a manifold has a boundary, the Gibbons-Hawking-York term has to be added to the action.
	\item {\it Palatini-Einstein formalism}: The action is defined by~\eqref{action:EH}. The connection and the metric are regarded as independent variables. Either the metricity or the torsion-free condition is imposed rather than both of them. Varying the action with respect to the connection, one learns that the connection to be the Levi-Civita connection. Thus, in the Palatini-Einstein formalism, either the metricity condition or the torsion-free condition is unnecessary. 
	\item {\it The formulation of this paper}: The action is defined by~\eqref{action:R+W}. The connection and the metric are regarded as independent variables. No conditions on the connection are imposed. The variation of the action with respect to the connection identically vanishes. Thus, in this formulation, both the metricity condition and the torsion-free condition are unnecessary. Even when a manifold has a boundary, it is unnecessary to add the Gibbons-Hawking-York term, because the action includes it. 
\end{enumerate}

\section{Examples\label{sec:examples}}
The $W^\lambda{}_{\mu\nu}$ and $W$ can be written in terms of the metric and the connection. Using \eqref{def:Levi-Civita} and \eqref{def:tensorW}, we obtain
\begin{align}
	W^\lambda{}_{\mu\nu}
	=&\frac{1}{2}\left(
	T^\lambda{}_{\mu\nu} 
	+ g^{\lambda\rho}g_{\mu\sigma}T^\sigma{}_{\rho\nu}
	+g^{\lambda\rho}g_{\nu\sigma}T^\sigma{}_{\rho\mu}
	\right)\nonumber\\
	+& \frac{1}{2}g^{\lambda\rho}\left(\nabla_\rho g_{\mu\nu} - \nabla_\mu g_{\nu\rho} - \nabla_\nu g_{\mu\rho}\right),
	\label{tensorW:TQ}
\end{align}
where $T^\lambda{}_{\mu\nu} \equiv \Gamma^\lambda{}_{\mu\nu} - \Gamma^\lambda{}_{\nu\mu}$ is a torsion tensor. Substituting \eqref{tensorW:TQ} into \eqref{def:scalarW}, we find that
\begin{align}
	W&=
	T^{\mu\nu}{}_\lambda Q^\lambda{}_{\mu\nu}
	+T^\lambda{}_{\lambda\nu} Q_\mu{}^{\mu\nu}
	-T^\lambda{}_{\lambda\nu} Q^{\nu\mu}{}_\mu
	\nonumber\\
	&-\frac{1}{4}T^\lambda{}_{\mu\nu} T_\lambda{}^{\mu\nu}
	+\frac{1}{2} T^{\mu\nu}{}_\lambda T^\lambda{}_{\mu\nu}
	+T^\lambda{}_{\lambda\mu} T_\nu{}^{\nu\mu}\nonumber\\
	&-\frac{1}{4} Q^{\mu\nu}{}_\lambda Q_{\mu\nu}{}^\lambda
	+\frac{1}{2} Q^{\mu\nu}{}_\lambda Q^\lambda{}_{\mu\nu}\nonumber\\
	&+\frac{1}{4} Q^{\mu\lambda}{}_\lambda Q_{\mu\nu}{}^\nu
	-\frac{1}{2} Q^{\mu\lambda}{}_\lambda Q^\nu{}_{\nu\mu},
	\label{scalarW:TQ}
\end{align}
where $Q_{\lambda\mu\nu} \equiv \nabla_\lambda g_{\mu\nu}$ is a non-metricity tensor. Then, substituting \eqref{tensorW:TQ} into \eqref{R+W:tensorW}, we obtain
\begin{align}
	R + W = \overline{R} 
	+ \overline{\nabla}_\mu \left(2T_\nu{}^{\mu\nu} +Q^{\mu\nu}{}_\nu - Q_\nu{}^{\mu\nu}\right),
	\label{R+W:TQ}
\end{align}
which is valid for any connection. 

General relativity is obtained as the simplest example by imposing two conditions on the connection,
\begin{align}
	&\mbox{metricity:} &Q_{\lambda\mu\nu}&\equiv\nabla_\lambda g_{\mu\nu} = 0,\label{def:metricity}\\
	&\mbox{no torsion:} &T^\lambda{}_{\mu\nu}&\equiv\Gamma^\lambda{}_{\mu\nu} - \Gamma^\lambda{}_{\nu\mu}= 0.\label{def:torsionfree}
\end{align}
Then, \eqref{tensorW:TQ} and \eqref{scalarW:TQ} reduce to $W^\lambda{}_{\mu\nu}=0$ and $W=0$. This means that the connection is uniquely determined as the Levi-Civita connection. Therefore, the action~\eqref{action:R+W} reduces to $S_{\rm GR}$, which is given by the first term in~\eqref{action:GR+boundary}.

If only the metricity condition~\eqref{def:metricity} is imposed rather than both \eqref{def:metricity} and \eqref{def:torsionfree}, then \eqref{tensorW:TQ} and \eqref{scalarW:TQ} reduce to 
\begin{align}
	W^\lambda{}_{\mu\nu}
	&=\frac{1}{2}\left(
	T^\lambda{}_{\mu\nu} 
	+ g^{\lambda\rho}g_{\mu\sigma}T^\sigma{}_{\rho\nu}
	+g^{\lambda\rho}g_{\nu\sigma}T^\sigma{}_{\rho\mu}
	\right),\\
	W &= 
	-\frac{1}{4}T^\lambda{}_{\mu\nu} T_\lambda{}^{\mu\nu}
	+\frac{1}{2} T^{\mu\nu}{}_\lambda T^\lambda{}_{\mu\nu}
	+T^\lambda{}_{\lambda\mu} T_\nu{}^{\nu\mu}.
	\label{def:scalarT}
\end{align}
In this case, the $W^\lambda{}_{\mu\nu}$ is called the contorsion tensor, and the $W$ is the torsion scalar usually denoted by~$T$. Then, \eqref{R+W:TQ} reduces to 
\begin{align}
	R + T &= \overline{R} + 2 \overline{\nabla}_\mu T_\nu{}^{\mu\nu},
	\label{R+T:tensorT}
\end{align}
where the Ricci scalar $R$ and the torsion scalar $T$ are in general nonzero.

Teleparallel gravity~\cite{Aldrovandi:2013wha} is a special case of~\eqref{R+T:tensorT}. Its theoretical structure, extensions, and physical applications have been studied in the literature~\cite{Li:2010cg,Sotiriou:2010mv,Pereira:2013qza,Krssak:2015oua,Golovnev:2017dox,Oshita:2017nhn,Krssak:2018ywd,Hohmann:2019nat,Pereira:2019woq}. It uses the Weitzenb\"ock connection~\cite{Weitzenbock:1923efa,Einstein:1928:RGA,Unzicker:2005in}, which leads to nonzero torsion but zero curvature~\cite{Einstein:1928:RGA}. Then, \eqref{R+T:tensorT} reduces to 
 \begin{align} 
	\widetilde{T} &= \overline{R} + 2 \overline{\nabla}_\mu \widetilde{T}_\nu{}^{\mu\nu},
	\label{torsionscalar:teleparallel}
\end{align}
where a tilde is used to denote the quantities for the Weitzenb\"ock connection. Eq.~\eqref{torsionscalar:teleparallel} defines teleparallel gravity, and hence teleparallel gravity is the case of \eqref{R+T:tensorT}.

If only torsion-free condition \eqref{def:torsionfree} is imposed rather than both~\eqref{def:metricity} and~\eqref{def:torsionfree}, then \eqref{tensorW:TQ} and \eqref{scalarW:TQ} reduce to
\begin{align}
	W^\lambda{}_{\mu\nu}
	=&
	\frac{1}{2}g^{\lambda\rho}
	\left(Q_{\rho\mu\nu} - Q_{\mu\nu\rho} - Q_{\nu\mu\rho}\right),\\
	W=&
	-\frac{1}{4} Q^{\mu\nu}{}_\lambda Q_{\mu\nu}{}^\lambda
	+\frac{1}{2} Q^{\mu\nu}{}_\lambda Q^\lambda{}_{\mu\nu}\nonumber\\
	&+\frac{1}{4} Q^{\mu\lambda}{}_\lambda Q_{\mu\nu}{}^\nu
	-\frac{1}{2} Q^{\mu\lambda}{}_\lambda Q^\nu{}_{\nu\mu}.
	\label{def:scalarQ}
\end{align}
In this case, the $W^\lambda{}_{\mu\nu}$ is called the disformation tensor, and the $W$ is the non-metricity scalar usually denoted by~$Q$. Then, \eqref{R+W:TQ} reduces to 
 \begin{align}
 	R + Q &= \overline{R} + \overline{\nabla}_\mu \left(Q^{\mu\nu}{}_\nu - Q_\nu{}^{\mu\nu}\right),
	\label{R+Q:tensorQ}
\end{align}
where the Ricci scalar $R$ and a non-metricity scalar $Q$ are in general nonzero. Furthermore, if zero curvature is assumed, then only the non-metricity scalar $Q$ is nonzero, and that case is the symmetric teleparallel gravity~\cite{Nester:1998mp}. Its extension and physical applications have been studied~\cite{BeltranJimenez:2017tkd,Jarv:2018bgs,BeltranJimenez:2019tjy,Jimenez:2019ovq}. Thus, by imposing some conditions on the connection, the action \eqref{action:R+W} yields various cases. In this sense, the action \eqref{action:R+W} is regarded as a unified description. 

\section{Conclusion\label{sec:conclusion}}
The gravity action~\eqref{action:R+W} has been presented by using the connection-independence in gravity. The action contains the scalar $W$ in addition to the Ricci scalar $R$. The scalar $W$ can be written in terms of the torsion tensor and the non-metricity tensor, as shown in~\eqref{scalarW:TQ}. In this formulation, no conditions on the connection are imposed. Nevertheless, the action yields the Einstein equations. It is not necessary to add the Gibbons-Hawking-York term even when a manifold has a boundary, because the action includes it from the beginning. In this formulation, the dynamics is independent of a choice of connection. Therefore, any connection can be used if necessary, and hence the action~\eqref{action:R+W} yields general relativity, teleparallel gravity, symmetric teleparallel gravity, and others.\\

\noindent 
{\it Note added:} It has been realized that a similar construction had been studied in~\cite{Jimenez:2019ghw} from a different perspective.

\bibliography{references}

\end{document}